# Superconductivity: coherent "tunnelling" by a dielectric array of charge-carriers

Johan F. Prins

Department of Physics, University of Pretoria, Gauteng, South Africa

**Abstract**

It is accepted in the scientific literature that when superconduction manifests, a steady-state current flows through a material without an electric field being present; i.e. quite simply said: according to Ohm's law the electric field has to be zero when the resistance is zero. Theoretical models of superconduction ascribe this behaviour to the absence of scattering of the charge-carriers after they have formed a minimum energy configuration called a Bose-Einstein Condensate. It is argued here that the absence of scattering within a material, although absolutely necessary, is not sufficient to explain why the electric field between two contacts can become zero in order to register zero resistance. It is concluded that an electric field, and thus a resistance, must manifest unless (i) the charge-carriers form part of an array of dielectric centres and (ii) the charge-carriers can increase their velocities without increasing their kinetic energies. A model is presented which allows these seemingly contradictory requirements to manifest. The model is fitted to selected experimental results which have been published for low-temperature metals, YBCO and highly-doped p-type diamond. In each case a satisfactory description of the experimental results is demonstrated.

**PACS code:** 74.20.-z

**Keywords:** superconduction, mechanism for superconduction

**Postal Address:** P O Box 1537, Cresta 2118, Gauteng, South Africa
**Telephone:** +27 11 477-8005
**Facsimile:** +27 11 477-3709
**e-mail:** johanprins@cathodixx.com

# 1. Introduction

Generation of superconduction at high temperatures is one of the holy grails of physics. Tremendous excitement ensued after the discovery of superconducting ceramics during the 1980's [1]. At first there had been rapid progress in increasing the critical temperatures of these materials, but eventually a ceiling was reached at a critical temperature of ≈130 K. One of the problems hampering progress is that the mechanism responsible for the latter superconduction is not known. The BCS theory [2], which has been used since 1957 to model superconduction in the low temperature metals, could to date not model the properties of the ceramic superconductors. It is, however, still believed that two basic postulates, which can be derived from the earlier work of Ginsberg and Landau [3], and on which the BCS theory is also based, **must** still apply; i.e.

(i) The charge-carriers must be bosons

(ii) These charge-carriers (at a suitable critical temperature) form a Bose-Einstein Condensate; i.e. all the charge-carriers relax to the "same" lowest energy and therefore they cannot scatter when moving through the superconducting material.

In the case of the low-temperature metals the BCS approach models the charge-carriers as Cooper pairs. According to perturbation analysis, which is similar to those used in quantum field theory, a Cooper pair consists of two electrons which are bound together owing to the exchange of virtual phonons; this action supposedly "glues" the electrons to form pairs which are then able to act as bosons. The general consensus which is emerging is that the "glue" that is responsible for forming boson charge-carriers in the high-temperature ceramics might not be phonon-related. Not knowing the actual mechanism for these materials has reduced the search for higher temperature superconductors to alchemy!

Over the past decade claims for superconduction at temperatures higher than 130 K have been made on a fairly regular basis. Researchers in this field have become sceptical of such claims and insist on more information than a mere drop in the measured resistance to "zero". The reason for this is that there is no experimental measurement possible that can unequivocally prove that the electric potential between two contacts (to such a material) is "zero, definitely zero, and nothing else but zero". Without such a proof, one cannot claim unequivocally that the material is really a superconductor: in simple terms; according to Ohm's law a material with zero resistance **must** have no potential difference between two contacts when a current flows from one contact to another. In fact, this is the only characteristic that defines a superconductor unequivocally. But, since there is no voltmeter that is accurate enough to convincingly determine zero potential, it has become customary to make additional measurements of related properties. The two major effects used, and which are believed to confirm that an electric field is not required to drive such a current, are (i) the establishment of non-dissipating circular currents when switching on a magnetic field over the material and (ii) demonstration that the Meissner-Ochsenfeld effect manifests. It is argued here that although these measurements do prove that the charge-carriers are not scattering within the material (as required by the BCS model); it does not, however, eliminate the possibility that these currents have been established by the presence of an electric field. Thus, they do not unequivocally prove that there will not be an electric field between two contacts to the material. Additional conditions have to apply for an electric field **not** to be present between two contacts. These conditions are postulated and used to model a general mechanism for superconduction that seems to explain all superconductors through which a current can flow without an electric field being present under all configurations possible.

**2. Traditional magnetic verification of superconduction: circular currents**

*2.1 Non-dissipating circular currents*
When switching on a magnetic field over a conducting material, circular currents are induced during the time that the magnetic field increases from zero to its maximum value. This is in accord with Faraday's law of induction according to which a circular electric field is generated around a changing magnetic field. It is accepted in the standard literature on superconduction that this electric field is responsible for driving the charge-carriers in order to establish the induced circular currents. This implies that the charge-carriers do experience the induced electric field and thus increase their velocities; i.e. they are accelerated. It is known experimentally that superconducting charge-carriers experience transient electric fields. One might want to argue that this is the expected reaction of the charge-carriers because the electric field is a transient field which disappears when the magnetic field has reached its maximum value. The charge-carriers then keep on moving in circles without an electric field being present; and this proves that superconduction is taking place.

A subtle problem, however, remains. When switching on the magnetic field one can, for example, increase it at a constant rate over a long period of time. The induced electric field will then be constant during most of this time. According to Lenz's law the induced electric field must keep on increasing the circular currents so that a magnetic field is generated that opposes the increasing applied magnetic field. This can only mean that the charge-carriers are being "accelerated" by the induced "non-transient" electric field. Thus, although the generation of the circular currents do prove that they will not dissipate when there is no electric field, it does not prove that the charge-carriers cannot be accelerated by an electric field, static or transient, when such a field is present. In other words, the absence of the electric field might not be a result of superconduction; because the field will also be absent in a normal conductor. Furthermore, not all the charge-carriers are participating in the formation of these currents; usually only those near the surface. Does this not imply that these carriers have higher energies than the rest of the carriers within the superconducting phase? Does this not make nonsense of the postulate that the charge-carriers are all in the "same" minimum energy state as required for the formation of a Bose-Einstein Condensate?

*2.2 Meissner-Ochsenfeld effect*

In the case of the Meissner-Ochsenfeld effect, a static magnetic field is applied while the material is not superconducting. When then cooling the material to become superconducting it is found that the magnetic field is expelled from the material. The reason for this is that circular currents appear near the surface of the material and they then generate a magnetic field that is equal to and opposite to the applied magnetic field. Again, after equilibrium has been reached, there is no electric field driving the currents; i.e. superconduction is taking place. But does this effect prove that the charge-carriers will not experience an electric field when one is present?

Consider the material when it is not yet superconducting. When a charge-carrier moves in a direction perpendicular to the applied magnetic field it will be deflected in order to follow a circular trajectory around this direction. The Lorentz force acting on the charge-carrier is in reality an electric force induced by the movement of the charge-carrier relative to the magnetic field. In a resistive material the charge-carrier will scatter from following a circular trajectory or repeated circular trajectories. The weaker the scattering becomes, the more easily it becomes to complete a circular trajectory. If the scattering becomes zero it should keep on performing circular motion. More and more charge-carriers will then be induced to participate in this movement, until the magnetic field they generate cancels the applied magnetic field. Any induced electric field then again falls to zero. Thus the Meissner-Ochsenfeld effect also does not prove beyond reasonable doubt that the charge-carriers will not be accelerated by an electric field when such a field is applied. Furthermore, again not all charge-carriers are participating to form the currents; so again the charge-carriers are not all at the "same" minimum energy as required for a Bose-Einstein Condensate to manifest.

**3. Currents flowing between two contacts through a superconductor**

*3.1 Electric field and charge-carrier velocity*

It has just now been concluded that when circular currents are generated by a changing magnetic field, the induced electric field could and most probably does accelerate the charge-carriers. So when switching on an electric field between two contacts it should also accelerate the charge-carries. In fact, it can be experimentally verified that the velocities of the charge-carriers do increase when the EMF around the circuit, within which the superconductor is an element, is switched on. This is strong proof that the charge-carriers within the superconducting element must be accelerated; however, the velocities eventually reach a steady-state value which is related to the magnitude of the EMF. This is exactly what happens when a resistance manifests! If the material is truly superconducting this behaviour must imply that an electric field is experienced by the charge-carriers between the contacts until they reach a critical velocity, at which point the electric field becomes identically zero. How is this possible?

The mechanism which causes the electric field to go to zero when current flow is induced by a changing magnetic field, cannot manifest between two contacts. Yet, also in the latter case, the electric field must fall to zero once the applicable carrier-velocity has been reached, or else the material cannot be a superconductor. So **why** would there be no electric field once an equilibrium-current manifests between two contacts? If the electric field is zero, it must imply that there must be an inbuilt mechanism that cancels the electric field when the charge-carriers reach a critical velocity. It is eminently clear that this mechanism cannot be the same as the one for circular currents induced by a magnetic field. The absence of an electric field between two contacts can only be explained if the superconductor acts as a perfect dielectric while it conducts a critical equilibrium-current between two contacts.

What constitutes a perfect dielectric? When applying an electric field across a perfect metal by placing it between two capacitor plates the metal acts as a perfect dielectric; i.e. it totally cancels the field within itself. A zero electric field is, however, not possible when applying the field via two contacts so that a current flows. In fact, the current flows in an attempt to cancel the field within the metal; but does not succeed to do so because the charge-carriers are continuously being replenished at the injecting contact and extracted at the end contact. How does a superconductor then manage to cancel an electric field (applied by contacts) while a current is flowing at the same time? Nobody has ever given a mechanism for this aspect which, I postulate, is the most important aspect that needs to be explained in order to understand how superconduction manifests.

*3.2 Charge-carrier velocity and kinetic energy*

It is clear that when the EMF is increased around a circuit containing a superconducting element, the charge-carriers are "accelerated" to a higher velocity within the superconductor. Does this not mean that their kinetic energies increase above the minimum energy they require to form a Bose-Einstein Condensate? Furthermore, an increase in their kinetic energies must be dissipated at some stage. Consider, for example, a current that flows between the cathode and anode in a vacuum diode. These electrons do not scatter while they move from one contact to the other. Does this mean that they are forming a superconducting phase? No not at all! The applied electric field accelerates them so that their kinetic energy increases. Accordingly, they have to shed this energy by scattering within the anode. Heat is still generated; i.e. a resistance manifests. Thus if the kinetic energy of the charge-carriers in a superconductor increases when increasing the EMF, one would expect these charge-carriers to similarly scatter within the contact they move into. Resistance will then manifest.

A possible way to try and wriggle out of this dilemma would be to argue that the velocity of the charge-carriers is determined by the drift velocity of the charge-carriers within the contacts to the superconductor. Thus, when charge-carriers are injected at the one contact, they enter with the same drift velocity they (just) had within this contact, and this determines the "drift velocity" within the superconductor until they enter the other contact within which they have the same drift velocity. Assume, however, that the two contacts to the superconductor are made of different metals which have different densities of charge-carriers. Then the drift velocities within the contacts will be different when the same current flows through them. If, for example, the drift velocity within the end contact is less than within the injecting contact, the superconducting charge-carriers will be injected into the superconducting material with a higher drift velocity than required within the end contact. When reaching and entering the latter contact their kinetic energy will be higher than required. So they **will have** to scatter within this contact in order to lose energy. This will be registered as resistance.

**4. What happens to the applied electric field between two contacts?**

*4.1 A summary*
The salient conclusions drawn from the arguments above can be summarised as follows:
1. There is not a single experiment known at present that can be, or has been, used to conclusively demonstrate that an electric field is not required to generate and maintain a current between two contacts to a superconductor.
2. To fully explain superconduction between two contacts without an electric field being present one needs:

    (i) A mechanism that explains how a current can flow through a superconductor while this material simultaneously acts as a perfect dielectric.
    (ii) A mechanism that explains why the charge-carriers can enter a contact with a high velocity without scattering.

Is current flow between two contacts possible without an electric field being present? It is clear that this has not yet been unequivocally demonstrated by **any** experiment to date. So we really do not know. Magnetic experiments only prove that circular currents can flow when there is no electric field present once the applied magnetic field reaches a constant value. As logically argued above, the same argument cannot be used when an electric field is directly applied to manifest between two contacts; i.e. in this case the electric field cannot go to zero in the same way. Assuming, however, that such a superconducting phase can exist between two contacts, how can the conundrums, listed under point 2, be explained? The perfect dielectric required cannot be a perfect metal. The only other type of dielectric known is a material in which dipoles are generated throughout the material when applying an electric field over the material. Thus to model point 2(i), one is forced to postulate that each charge-carrier must form part of a dielectric array so that it can be polarised relative to other opposite charges when an electric field is applied. Point 2(ii) can only be explained by making another contentious assumption, which, in turn, raises another conundrum: this is to assume that although the charge-carriers within a superconductor have velocities, and can increase their velocities, these velocities do not manifest as kinetic energy which requires dissipation within the contact that the charge-carriers are moving into.

*4.2 Dielectric properties of a superconductor*

If the charge-carriers can be polarised relative to their opposite charges by an applied electric field, it implies that they must (on average) be stationary when no current is flowing. If the charge of the carriers is, for example, negative, they must be "bonded" to equal positive charges; i.e. they can either be part of a permanent array of dipoles, or form part of an array of induced dipoles when an external electric field is applied. In either case, such an applied electric field must polarise the charge-carriers relative to the positive charges that "anchor" them. Such increased polarisation of the charge-carriers must increase their energies. If, without polarisation a charge-carrier requires an increase in energy $\Delta E_C$ to break away from its anchor-point, then with polarisation it requires less energy to break away. If there is no applied electric field, a charge-carrier will have to acquire more than the energy $\Delta E_C$ to break free of its anchor-point and move to another, adjacent position where it can again become anchored. What is the required amount of energy? If the charge-carrier has a mass m and it moves with a speed v from its anchor-point to an adjacent position, it requires an amount of energy $\Delta E$ to break free of its anchor-point and reach the adjacent position which is given by:

$$\Delta E = \Delta E_C + \frac{1}{2}mv^2 \qquad (1)$$

Once it reaches the adjacent position it will have to get rid of this energy if it wants to again become anchored.

*4.3 Charge-carrier movement without generating kinetic energy*

In the standard scientific literature it is argued that a "particle" can "tunnel through" an energy barrier. Another (and, I believe, the scientifically correct) way to model "tunnelling", is to postulate that the "particle" can "borrow energy" for a short time interval $\Delta \tau$ in order to scale the energy barrier. This is made possible by Heisenberg's Uncertainty Relationship for energy and time, which is written as:

$$\Delta E \Delta \tau = g\eta \qquad g \geq 1 \qquad (2)$$

The value of g is determined by the wave function representing the charge-carrier. If the two anchor-points are separated by a distance R, and the charge-carrier "tunnels" with a speed v from one to the other, then in order not to violate Heisenberg's Uncertainty Relationship, one must have that:

$$v = \frac{R}{\Delta \tau} \tag{3}$$

Combining Eqs. 1, 2, and 3 leads to:

$$v^2 - \left(\frac{g\eta}{mR}\right)v + \left(\frac{2\Delta E_C}{m}\right) = 0 \tag{4}$$

This equation can be readily solved to obtain:

$$v = \frac{g\eta}{mR}\left(1 - \sqrt{1 - \frac{2m\Delta E_C R^2}{(g\eta)^2}}\right) \tag{5}$$

One can immediately see that when R increases, a maximum value R(max) will be reached above which the solution for v becomes a complex number; i.e. there exists a maximum velocity v(max) for v. When solving for R(max) and v(max) one obtains that:

$$R(\max) = \frac{g\eta}{\sqrt{2m\Delta E_C}} \tag{6}$$

and

$$v(\max) = \frac{g\eta}{mR(\max)} = \sqrt{\frac{2\Delta E_C}{m}} \tag{7}$$

Both parameters, surprisingly, turn out to only depend on the binding energy ($\Delta E_C$). Accordingly, even for distances R<R(max) the maximum possible "tunnelling" speed is still given by Eq. 7.

*4.4 A mechanism for current flow and cancelling the applied electric field at the same time*
It is now postulated that to be able to superconduct, a material must contain an array of stationary charge-carriers, which can be polarised by an externally applied electric field. How such arrays can form, and examples of how to calculate their physical properties have been published elsewhere [4]. When the density of such an array becomes high enough, so that the distances R between the charge-carriers along a direction within the material become smaller than R(max), the charge-carriers can move from anchor-site to anchor-site by borrowing energy as described in section 4.3. When only applying an external electric field the charge-carriers become polarised. This will happen very quickly because the array of charge-carriers will act as a dielectric so that the electric field will spread out over them with the same speed that is possible for light within such a material. When the electric field is applied via contacts, then simultaneous injection of new charge-carriers could occur. The original charge-carries will still become aware of the electric field spreading through the material at the speed of light, however, they will not become polarised because this will increase their energies. They are able to maintain the same energies they have had before the electric field has been switched on by "tunnelling" from their anchor-points to adjacent ones; i.e. the charge-carriers which are injected at the contact, can displace existing charge-carriers from their anchor-points, which in turn can displace other charge-carries from their anchor-points etc. Charge-carrier movement thus occurs in order to maintain the lowest energy state as is required by the second law of thermodynamics.

The important point to notice is that although each charge-carrier has an uncertainty in its energy during its movement from one anchor-point to the next equivalent anchor-point, it does not permanently gain energy; i.e. the energy of a charge-carrier arriving at an anchor-point is the same as the energy it has had at the previous anchor-point. The charge-carriers are thus not gaining energy and therefore they cannot be scattered when they reach and enter an end-contact. Furthermore, when such a current flows, the charge-carriers are marching in step from the one contact to the next. Just as many charge-carriers are ejected at the end contact as those that are injected at the initial contact. Thus, on average, the charge-carrier density stays the same as it has been before the current started to flow. Therefore, the carrier-charges and anchor-charges still act together to manifest a perfect (but now dynamic) dielectric. They cancel the applied field. "Real" superconduction occurs. One might thus define superconduction as coherent "tunnelling" by a dielectric array of charge-carriers; however, I believe that the term "tunnelling" could be a misleading term. "Tunnelling" implies a "particle" moving "through" a barrier. Here the process is described by charge-carriers borrowing energy to scale barriers. Nonetheless, I will call the process coherent "tunnelling" for the time being.

It should be clear that when increasing the applied electric field this will be "felt" by all the carriers at the speed of light, and they will move faster in order to accommodate the increases in injected carriers. This increase in speed, however, still does not manifest as an increase in kinetic energy.

**5. Explaining the properties of superconductors**

*5.1 Maximum currents*
It is a well-known experimental fact that a superconductor has a maximum electrical current that can flow through it. If one tries to increase the current above this value, it falls away. According to the mechanism postulated above (paragraph 4.4), this behaviour becomes logically consistent. When increasing the field, the "drift" speed v increases until it reaches the maximum value possible v(max) (see Eq. 7). Any further increase leads to non-real speeds and thus to an impossible current.

*5.2 Temperature behaviour*
Each charge-carrier has a binding energy ($\Delta E_C$). Thus its bonding can be thermally destroyed: i.e. the charge-carrier can be excited to a higher energy level. This also causes it to neutralise its positive anchor charges [4]. If the density of charge-carriers at absolute zero temperature would be $N_0$, then the density of charge-carriers N at temperature T can be written as:

$$N = N_0 \left(1 - \exp\left[-\frac{\Delta E_c}{2k_B T}\right]\right) \quad (8)$$

Thus, as the temperature increases, the density of charge-carriers N decreases. This means that the average distance R (between the carriers) increases. When this distance becomes larger than the maximum distance R(max) (see Eq. 6), coherent tunnelling is not possible anymore, and the superconducting current thus disappears. The critical temperature $T_c$ has been reached. Since (according to the present model) superconduction is considered along a single direction, it is the density of charge carriers along this direction that determines the process. Eq. 8 can thus be written in terms of the concomitant "tunnelling" distances as:

$$R = \frac{R_0}{1 - \exp\left[-\frac{\Delta E_c}{2k_B T}\right]} \quad (9)$$

*5.3 Non-dissipating circular currents revisited*
When switching on a magnetic field, the induced circular electric field will attempt to polarise the charge-carriers relative to their anchor-charges; however, they avoid polarisation by rather tunnelling along circular paths in order to avoid an increase in energy. As already discussed above (for current flow between two contacts) the circular currents manifest in order to maintain a minimum-energy dynamic-equilibrium state as required by the second law of thermodynamics. Since this is a minimum-energy state, the charge-carriers cannot, for example, radiate electromagnetic waves.

*5.4 Meissner-Ochsenfeld effect revisited*
When applying a magnetic field, the energy of the charge-carriers also increases. This can be demonstrated by solving the Schrödinger equation subject to using the dynamical momentum operator **p** [4]; i.e.

$$\mathbf{p} = -i\eta\nabla - (-e)\mathbf{A} \quad (10)$$

This will not be done here because the solution is different for different materials. It should, however, be clear that here, just as in the case of the induced circular currents (paragraph 5.3), the circular currents form so that the charge-carriers do not have to increase their energies, and so that the whole system can relax to the lowest-energy dynamic-equilibrium state, as required by the second law of thermodynamics. Thus, also in this case electromagnetic radiation is not possible.

*5.5 Magnetic strength of a superconductor*

When applying a magnetic field, currents evolve to expel the magnetic field, so that the charge-carriers can maintain their lowest energy state (paragraph 5.4). As already deduced above (paragraph 5.1), such a current cannot be sustained if the magnitude of the velocity required to sustain the opposite magnetic field approaches the maximum possible speed v(max). Thus, once the applied magnetic field becomes so large that the latter condition would be required, the charge-carriers are unable to tunnel fast enough; the superconducting state can thus not manifest. At $T=T_c$ the distances between the charge-carriers are equal to R(max). The charge-carriers have to tunnel at the maximum speed v(max); i.e. they are moving at their limit even without a magnetic field being present. When now applying a magnetic field, a higher speed is needed to counteract the applied magnetic field: this is not possible. At lower temperatures $T<T_c$, there are more charge-carriers, so that the distances between them are smaller than R(max); i.e. they can transport a current at a lower speed than v(max). When now applying a magnetic field, the charge-carriers can move in order to manifest a current which, in turn, forms an opposite magnetic field that cancels the applied field. This is possible until the magnitude of the applied field becomes so large that the magnitude required, for the velocities of the charge-carriers, again becomes too large. Accordingly, as the temperature decreases the critical magnetic field required to prevent a superconducting phase from forming will keep on increasing: This is exactly as experimentally observed.

*5.6 Josephson tunnelling*

When Josephson made the theoretical prediction that superconducting charge-carriers should be able to tunnel through a thin insulator [5], John Bardeen rejected this possibility because (Cooper) "pairing does not extend into the barrier, so there can be no such superfluid flow". Nonetheless Josephson tunnelling was subsequently demonstrated by experiment which indicates that Bardeen had been wrong; however, has he been wrong? Does Josephson tunnelling not rather prove that the Cooper mechanism is not tenable?

Josephson tunnelling is a logical consequence of the mechanism postulated here. The charge-carriers transport current by tunnelling from anchor point to anchor point, so they should also be able to tunnel "through" a thin insulator. Furthermore, when the distance through which the latter tunnelling occurs is different from that within the superconductor, an extra phase shift should result. This can be modelled by using simple periodic boundary conditions. This will not be pursued any further here.

**6. Applying the model to experimental results**

*6.1 Low-temperature metal superconductors*

According to the model postulated here, superconducting charge-carriers must form part of a dielectric array (see paragraph 4.4.). Is this possible within a metal? According to Eugene Wigner it is possible [6].

Electrons can form a "Wigner crystal" within a "non-ideal" metal. Wigner analysed the validity of the mean field approach which is used to model electrons in a metal. According to this approach each electron is fully screened by the other electrons so that they do not experience the positive charges on the metal ions. He found that for "non-ideal" metals, a metal-insulator transition should occur (at a low enough temperature) through the formation of an array of electron-related pseudo-particles (which will be referred to as orbitals) which are stationary. This occurs because the electrons now become aware of the positive charges and therefore they relax to a lower energy in order to form such an array. The orbitals are stationary because each of them is anchored by the positive charges that form when the concomitant electrons de-excite from their mean-field energy states. There is, of course no reason why these orbitals could not be bi-electron orbitals. In fact they most probably will be, because such an array will have a lower energy. Wigner also deduced that each orbital should radially be a Gaussian-shaped wave function.

Once all the mean-free electrons have lost energy to form orbitals. The electrons are all localised at specific lattice sites; therefore the term "Wigner crystal". For this reason the metal should then become an insulator; however, according to the model proposed here, superconduction should be possible along any direction if the distances R (between the orbitals) become short enough. This implies that the charge-carriers are formed by a quantum mechanical mechanism that is similar to the one responsible for the formation of stationary bi-electron orbitals around the nucleus of the atom. No "virtual" phonon interaction is involved to "bind" the electrons together. It also immediately explains why high quality conductors like gold and copper do not display superconduction; Wigner crystals cannot form within such materials.

The experimental points shown in Fig. 1 have been measured by Townsend and Sutton [7]. A thin insulating layer was sandwiched between the superconductor and a normal metal. The insulator was made thin enough so that single electrons can tunnel through it. An electric potential $\Phi$ was then applied over the insulating layer. Charge-carriers within the superconductor that tunnelled into the insulating layer experienced the resultant field within the layer and they accelerated. According to the BCS model, the two electrons constituting each superconducting charge-carrier, are bonded together by the exchange of virtual phonons in order to form a Cooper pair, and the resultant binding energy is given by $\Delta E_{SC}$. Thus, as soon as a charge-carrier has accelerated so that each of its electrons has gained energy of $\frac{1}{2}\Delta E_{SC}$, it breaks up into two separate entities. By measuring the potentials, $\Phi_S$, at which single electrons reach the opposite side of the insulator, one can calculate the binding energy from $\frac{1}{2}\Delta E_{SC} = e\Phi_S$. According to the BCS theory the increase in $\Delta E_{SC}$ with decreasing temperature relates to an increase in the effectiveness of electron-phonon coupling.

According to the model espoused here, the two electrons are bonded by a constant binding energy equal to $\Delta E_C$. When such an orbital tunnels into the insulator with a speed v it is accelerated to increase its speed further until the total kinetic energy equals the binding energy. At this point the bi-electron charge-carrier split into two separate entities, which can tunnel further on their own. The amount of extra energy needed after the charge-carrier enters the insulator is thus $\Delta E_C - \frac{1}{2}mv^2$, and this is equal to $\Delta E_{SC}$; i.e. the following equation describes the process:

$$\Delta E_{SC} = \Delta E_C - \frac{1}{2}mv^2 = 2e\Phi_S \qquad (11)$$

The mass m has been taken as equal to twice the electron mass (since the orbital is a pseudo-particle the mass could be different). As the temperature decreases, the speed of tunnelling v also decreases so that $\Delta E_{SC}$ increases. The theoretical curves in Fig. 1 have been fitted according to the model postulated above in conjunction with Eq. 11; i.e. in conjunction with Eqs. 5 and 9. The binding energy for tin is found to be $\Delta E_c$ =1.4 meV and for tantalum $\Delta E_c$ =1.68 meV. The BCS theory does not render such high quality theoretical curves [7].

It is important to note that within a material the electronic levels are also functions of the phonon spectrum. For this reason it has become customary to talk of "vibronic" levels. Thus if the atoms are replaced by their isotopic counterparts, the orbital energy $\Delta E_C$ will change. So will then the critical temperature $T_c$. Thus, the isotope effect in the low temperature superconductors need not relate to phonon exchange between electrons at all. In fact, I believe that modelling superconductors in terms of Cooper pairs and getting some semblance of correspondence with experimental results have been fortuitous.

*6.2 Superconduction in heavily-doped p-type diamond*

It has been discovered that heavily-doped p-type diamond superconducts at low temperatures [8]. The dopant density is so high that the density of boron acceptors is on average above the threshold for the Mott transition; i.e. they form an impurity-band. Attempts have thus been made to model the mechanism along the rules of the BCS theory, but without any success. Experimental data points measured on such a diamond by Bustarret et al are shown in Fig. 2 [9].

When a semiconductor becomes highly doped so that impurity-band conduction ensues, not all of the dopant centres become delocalised. At the edges of the impurity-band there are tails of localised centres, below the so-called mobility edge of the impurity-band. Thus, as the width of the acceptor impurity-band increases with increased doping, localised acceptor states are forming at lower and lower energies relative to the valence band. When the temperature decreases to a low enough value, the delocalised acceptor states will depopulate and the lower energy localised states will be the only ones which still accommodate electrons. Eventually only the lowest-energy states will on average have electrons and they will then depopulate when the temperature drops further. At this stage the Fermi-level should energetically lie about half-way between the impurity band and the valence band. This is so because the impurity band now acts as a "conduction" band. Although the activation energy $\Delta E_H$ for hole-generation in the valence band has now become very small, the density of holes must still be proportional to the density of acceptors $N_A$; so that one can write that:

$$p = CN_A \exp\left[-\frac{\Delta E_H}{k_B T}\right] \qquad (12)$$

C is a proportionality constant which is a function of the valence band electron-state density and might be a weak function of temperature. By using this equation in conjunction with the equations above, and assuming that the binding energy $\Delta E_C$ is equal to $2\Delta E_H$ the curve in Fig. 2 has been fitted theoretically to the data points. The value obtained for the binding energy is $\Delta E_c = 0.41$ meV and the maximum distance at which superconduction sets is R(max) = 96.4 Å. This implies that the density of the shallow-energy acceptor levels responsible for the superconducting phase to just form at absolute zero must be $\approx 1.1 \times 10^{18}$ cm$^{-3}$; a very plausible result.

It is important to note that the charge-carriers have here been modelled as single holes; i.e. fermions. As the temperature increases more electrons are excited from the valence band, the density of holes decreases until the distances between them become larger that R(max); superconduction then ceases. The following question might now arise: why would the holes move to replace other holes when they are surrounded by acceptor centres populated by electrons? Should the electrons not tunnel into adjacent holes? They might do just that, however, even if the electrons do tunnel, the average density of holes does not change.

*6.3 Ceramic superconductors: YBCO*

YBCO is probably the best studied ceramic superconductor. Within the perfect YBCO orthorhombic structure, which itself constitutes a superconducting phase, the number of oxygen atoms per unit cell is equal to 7. YBCO is, like most of these materials, a layered structure. $CuO_2$ planes are separated by Y atoms. Below these planes there are BaO planes and then planes within which oxygen atoms are aligned along chains. In YBCO, as in many of the other related "high-temperature" superconducting materials, oxygen is usually stoichiometrically deficient so that the number of oxygen atoms per unit cell is less than 7. It is found experimentally that superconduction occurs as long as this number of oxygen atoms lies within the following limits $6.4 < y \leq 7$. When y reaches the value 6, it is believed that there are none or very few oxygen atoms forming oxygen chains. Superconduction relates to the presence of the latter atoms. This is demonstrated by the experimental data points in Fig. 3, which have been measured and reported by Segawa and Ando [10]. The critical temperature increases as the oxygen content y increases above the value 6.4. There is, however, a plateau with the critical temperature at $\approx 60$ K for values of y between 6.6 and 6.8.

It will now be postulated that the oxygen atoms within the chains are (part of) donor centres which supply electrons which form arrays of bi-electron orbitals. The most probable manner in which these arrays can form is between the crystallographic planes. The possible mechanism involved has been modelled and published elsewhere [4]. The orbitals are Gaussian-shaped waves parallel to the crystallographic planes and are anchored by positive charges that reside on donors within the crystallographic planes. In the case of YBCO the density $N_e$ of electrons, which are available to form the required charge-carrier arrays, can thus be equated to the stoichiometric density y of the oxen atoms by writing that:

$$N_e = C(y - y_0) \qquad (13)$$

C is a proportionality constant that also adjusts the units, and $y_0$ is the density of oxygen atoms which are not situated along the chains. By using this equation in conjunction with the equations derived above, the theoretical curves shown in Fig. 3 have been fitted to the experimental data points. The dotted curve is an average fit through all the data points. In turn, the solid curve has been fitted through all the data points with the highest critical temperatures. For both curves one finds that $y_0$ is not equal to 6. For the solid curve it is $y_0$=5.348. This indicates that not all the non-chain oxygen sites have been filled when the chains started to form. Therefore one can explain the 60 K plateau as follows: when increasing the oxygen atoms, they at first keep on adding mostly to the chain-sites while those atoms not in chain positions stayed approximately constant at $y_0$=5.348. The plateau initiates when some, if not all the oxygen atoms being added, start to fill non-chain positions, until $y_0$ becomes equal to 6. After this point is reached, all the additional oxygen has to fill chain-positions again and the data points increase to again fall on the solid curve. The decrease in $T_c$, when y is equal to 7, can be explained by a decrease in $\Delta E_C$ when the charge-carrier density increases beyond a certain value [4]. The critical distance between the orbitals at which superconduction sets in can be calculated from Eq. 7 and it is found to be R(max)≈7.2 Å. It is approximately twice the lattice spacing along the basal plane of YBCO; a very satisfactory result. The binding energy for a bi-electron charge-carrier is found to be $\Delta E_C \approx 37.6$ meV; also a very plausible result.

Since the orbitals form between the crystallographic planes, the orbitals cannot interact directly with the phonon-levels within the crystallographic layers; i.e. their energies are not vibronic [4]. Thus, one does not expect a strong isotope effect as is observed for the low temperature metals; this is just as observed experimentally. This also indicates that in $MgB_2$, which shows an isotope effect, superconduction occurs within the crystallographic layers even though this material has a layered structure.

## 7. Discussion and conclusion

The mechanism proposed for superconduction in this exposé explains the known properties of superconducting materials very well (see paragraph 5). It also models the measured properties of different types of superconductors, be it low temperature metals, "high-temperature" ceramics or semiconducting superconductors, very well (see paragraph 6). What is especially satisfying is that a single mechanism describes all these different materials.

The proposed mechanism is **not** based on the requirements that electrons **must** pair and the resultant boson charge-carriers **must** then form a Bose-Einstein Condensate. In fact, the superconducting phase that forms in highly-doped p-type diamond can be adequately modelled in terms of the coherent tunnelling of holes; which are fermions. Although one can argue that the charge-carriers are all in the same minimum energy state as required for a Bose-Einstein Condensate, the dielectric array which is required for superconduction to manifest can be understood in terms of an array of separate entities; i.e. localised states. The role that quantum mechanics plays is through Heisenberg's Uncertainty Relationship for energy and time. This allows the charge-carriers to "tunnel" so that they can increase their velocities without increasing their kinetic energies. Thus, a mere absence of scattering of the charge-carriers, although necessary, is not (on its own) sufficient to ensure that superconduction will occur.

**Figure titles:**

Fig. 1:   The change in the apparent energy gap $\Delta E_{SC}$ with absolute temperature for tantalum and tin in their superconducting state. The experimental points were measured by doing a tunnelling measurement during which the bi-electron charge-carriers break up into separate electrons [7]. According to BCS theory, $\Delta E_{SC}$ is the binding energy between the two electrons (forming a charge-carrier) and this bonding is caused by virtual phonon exchange between the electrons. In the present model the two electrons form a bi-electron orbital by being bonded to opposite (positive) charges so that they have a constant binding energy equal to $\Delta E_C$. The temperature relationship of $\Delta E_{SC}$ is an artefact of the tunnelling process; i.e. $\Delta E_{SC}$ does not represent the real binding energy of the two electrons.

Fig. 2: Superconducting critical temperature as a function of boron density in highly doped p-type diamond [9]. The curve has been fitted by using the mechanism for superconduction derived in this publication. The charge-carriers are holes; i.e. fermions.

Fig. 3: Theoretical curves fitted to YBCO data [10]. The dashed curve is the best average fit, while the solid curve has been fitted through the highest critical temperatures.

**Figures:**

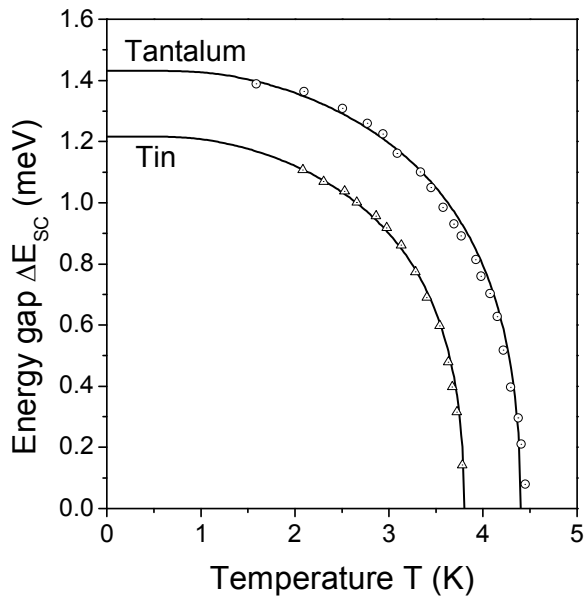

Figure 1

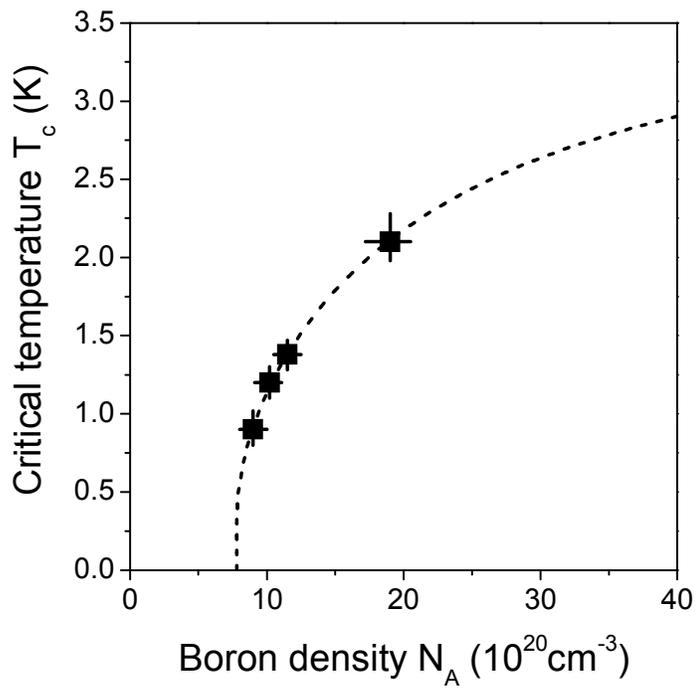

Figure 2

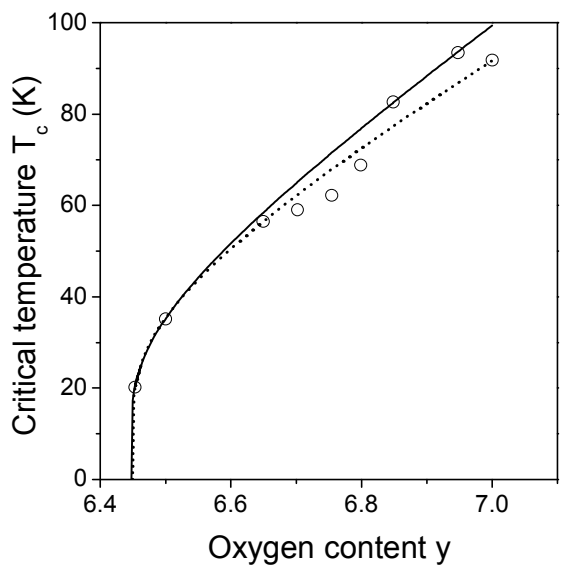

Figure 3